\documentclass[trackchanges,twocolumn]{aastex7}

\begin{document}

\title{The impact of organic hazes and graphite on the observation of CO\textsubscript{2}-rich sub-Neptune atmospheres }

\correspondingauthor{Chao He}
\email{chaohe23@ustc.edu.cn}

\author[0009-0009-2660-1764]{Haixin Li}
\affiliation{School of Earth and Space Sciences, University of Science and Technology of China, Hefei, Anhui 230026, China}
\email{}

\author[0000-0002-6694-0965]{Chao He}
\affiliation{School of Earth and Space Sciences, University of Science and Technology of China, Hefei, Anhui 230026, China}
\email{}

\author[]{Sai Wang}
\affiliation{School of Earth and Space Sciences, University of Science and Technology of China, Hefei, Anhui 230026, China}
\email{}

\author[]{Zhengbo Yang}
\affiliation{School of Earth and Space Sciences, University of Science and Technology of China, Hefei, Anhui 230026, China}
\email{}

\author[]{Yu Liu}
\affiliation{School of Earth and Space Sciences, University of Science and Technology of China, Hefei, Anhui 230026, China}
\email{}

\author[]{Yingjian Wang}
\affiliation{School of Earth and Space Sciences, University of Science and Technology of China, Hefei, Anhui 230026, China}
\email{}

\author[]{Xiao'ou Luo}
\affiliation{School of Earth and Space Sciences, University of Science and Technology of China, Hefei, Anhui 230026, China}
\email{}

\author[]{Sarah E. Moran}
\affiliation{NHFP Sagan Fellow, NASA Goddard Space Flight Center, Greenbelt, MD 20771, USA}
\email{}

\author[]{Cara Pesciotta}
\affiliation{Department of Earth and Planetary Sciences, Johns Hopkins University, Baltimore, MD, USA}
\email{}

\author[]{Sarah M. H{\"o}rst}
\affiliation{Department of Earth and Planetary Sciences, Johns Hopkins University, Baltimore, MD, USA}
\affiliation{Space Telescope Science Institute, Baltimore, MD, USA.}
\email{}

\author[]{Julianne I. Moses}
\affiliation{ Space Science Institute, Boulder, CO, USA}
\email{}

\author[]{V{\'e}ronique Vuitton}
\affiliation{Univ. Grenoble Alpes, CNRS, IPAG, 38000 Grenoble, France}
\email{}
\begin{abstract}
Many sub-Neptune and super-Earth exoplanets are expected to develop metal-enriched atmospheres due to atmospheric loss processes such as photoevaporation or core-powered mass loss. Thermochemical equilibrium calculations predict that at high metallicity and a temperature range of 300-700 K, CO${_2}$ becomes the dominant carbon species, and graphite may be the thermodynamically favored condensate under low-pressure conditions. Building on prior laboratory findings that such environments yield organic haze rather than graphite, we measured the transmittance spectra of organic haze analogues and graphite samples, and computed their optical constants across the measured wavelength range from 0.4 to 25 $\mu$m. The organic haze exhibits strong vibrational absorption bands, notably at 3.0, 4.5, and 6.0 $\mu$m, while graphite shows featureless broadband absorption. The derived optical constants of haze and graphite provide the first dataset for organic haze analogues formed in CO${_2}$-rich atmospheres and offer improved applicability over prior graphite data derived from bulk reflectance or ellipsometry. We implemented these optical constants into the Virga and PICASO cloud and radiative transfer models to simulate transit spectra for GJ 1214b. The synthetic spectra with organic hazes reproduce the muted spectral features in the NIR observed by Hubble and general trends observed by JWST for GJ 1214b, while graphite models yield flat spectra across the observed wavelengths. This suggests haze features may serve as observational markers of carbon-rich atmospheres, whereas graphite's opacity could lead to radius overestimation, offering a possible explanation for super-puff exoplanets. Our work supplies essential optical to infrared data for interpreting observations of CO${_2}$-rich exoplanet atmospheres.
\end{abstract}


\section{Introduction}\label{sec:Intro}
Super-Earths and sub-Neptunes are the most abundant exoplanet types discovered to date \citep[e.g.,][]{2011MayorMarmier+, 2012HowardMarcy+, 2013BatalhaRowe+, 2013FressinTorres+}. These planets, with radii between Earth and Neptune, have no direct analogs in the present-day Solar System, making them particularly important targets to improve our understanding of planetary atmospheres. A bimodal radius distribution—commonly referred to as the “radius valley”—suggests that many sub-Neptunes may evolve into super-Earths via their primordial hydrogen/helium envelopes loss through photoevaporation or core-powered mass loss \citep[e.g.,][]{2017FultonPetigura+, 2018VanEylenAgentoft+, 2019GuptaSchlichting+, 2021RogersOwen+}.

\begin{figure*}
    \centering
    \includegraphics[width=1\textwidth]{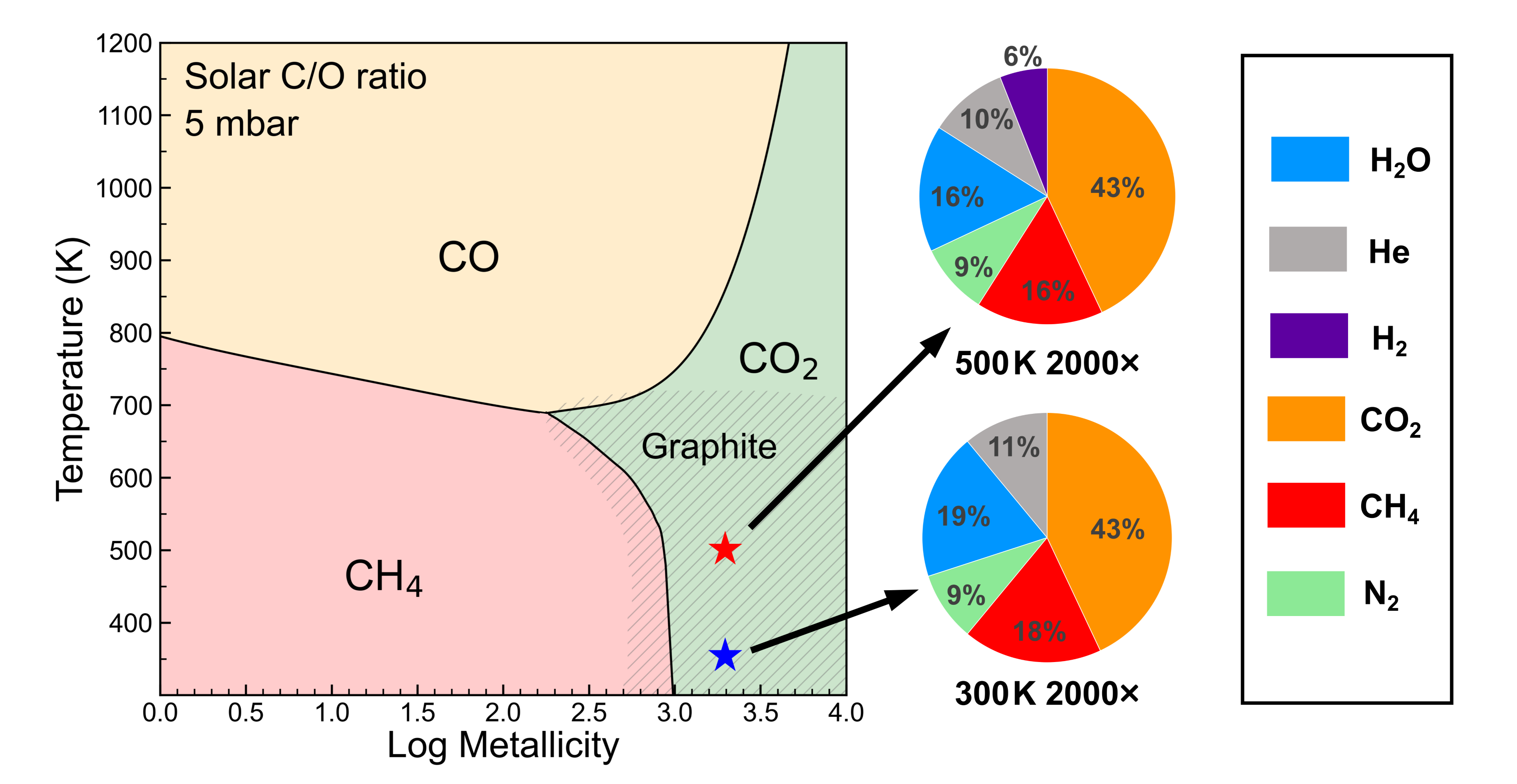}
    \caption{
    \textbf{Thermochemical equilibrium of carbon species.}
    The calculations from Chemical Equilibrium with Applications (CEA) code \citep{gordon1994computer} at 5~mbar and solar C/O ratio, $\sim$0.577 \citep{2009LoddersPalme+}, show the dominant carbon-bearing gas (CH$_4$, CO, or CO$_2$) as a function of temperature and metallicity. The shaded region indicates where solid graphite is thermodynamically favored. Red and blue stars mark the experimental conditions for the 500 and 300~K haze samples, both at 2000$\times$ solar metallicity. Pie charts illustrate the corresponding initial gas mixtures, composed of CO$_2$ (orange), CH$_4$ (red), N$_2$ (green), H$_2$O (blue), He (gray), and H$_2$ (purple), calculated from CEA without graphite in the list of species. The mean molecular mass of the 500 and 300 K cases are 27.4 g/mol and 28.2 g/mol, respectively. 
    }
    \label{fig:1}
\end{figure*}
The resulting stripped planets are expected to develop secondary atmospheres with high metallicity, often dominated by heavier molecules such as CO\textsubscript{2}. These atmospheres may enter a graphite-stable regime, where graphite is predicted to be the thermodynamically stable solid end product among carbon-bearing species \citep[e.g.,][]{2013MosesLine+, 2021WoitkeHerbort+}. Figure~\ref{fig:1} illustrates the dominant carbon species as a function of temperature and metallicity with solar C/O, $\sim$0.577 \citep{2009LoddersPalme+}, at 5 mbar as HST/JWST probe this region. Many observed sub-Neptunes and super-Earths have equilibrium temperatures of 300–800~K, overlapping with this regime. To investigate haze formation under such conditions, we simulate 2000$\times$ solar metallicity atmospheres at 300 and 500~K. In our companion study, we found that organic hazes rather than graphite form under such low-pressure CO\textsubscript{2}-rich conditions \citep{2025WangYang+}. However, graphite may still condense in deeper, hotter atmospheric layers and be uplifted into the upper atmosphere, motivating a comparative study of the spectral impacts of both organic haze and graphite aerosols.

Recent observational and laboratory evidence increasingly supports the ubiquity of haze in sub-Neptune and super-Earth atmospheres \citep[e.g.,][]{2014KreidbergBean+, 2018HeHorst+, 2018HorstHe+, 2024OhnoSchlawin+, 2024WallackBatalha+, 2025TeskeBatalha+}. Hazes can alter a planet’s thermal structure, albedo, and circulation, and suppress spectral features, complicating atmospheric characterization and habitability assessments \citep[e.g.,][]{2015MorleyFortney+, 2021GaoWakeford+}. Their impacts are strongly dependent on the optical properties, which vary greatly among different haze types (e.g., organics, soot, sulfur) and even within organic hazes formed under different conditions \citep{2015MorleyFortney+, 2017GaoMarley+, 2023CorralesGavilan+, 2024HeRadke+}. The atmospheric retrievals of exoplanet observations from advanced telescopes like JWST and the upcoming ARIEL mission \citep[e.g.,][]{2014BeichmanBenneke+,2018TinettiDrossart+} also strongly rely on the optical properties of laboratory-derived hazes. However, only a limited set of laboratory-derived haze reference data are available, mostly derived from N\textsubscript{2}- or H\textsubscript{2}O-rich haze analogs \citep[e.g.,][]{1984KhareSagan+, 2023CorralesGavilan+, 2024DrantGarcia-Caurel+, 2024HeRadke+}, leaving CO\textsubscript{2}-rich hazes poorly constrained. Using mismatched haze data may lead to inaccurate retrievals of key molecular abundances (e.g., H\textsubscript{2}O, CH\textsubscript{4}), as well as particle size, potentially leading to broader misinterpretations of atmospheric properties.

Meanwhile, the optical constants of graphite remain uncertain, with large discrepancies across datasets due to variations in sample morphology and measurement methods \citep[e.g.,][]{2000DjurisicLi+, 2003Drainea+, 2003Draineb+, 2008KuzmenkovanHeumen+, 2014PapoularPapoular+, 2016Draine+}. Most existing data are derived from bulk solids or thin films, which do not accurately represent suspended atmospheric aerosols, limiting their utility to atmospheric modeling and retrievals. Therefore, new laboratory measurements under conditions relevant to planetary observations are necessary. In addition, measuring optical properties of organic hazes and graphite using the same method enables the direct comparison of their spectral impacts and provides a more reliable dataset for modeling the potential spectral signatures of carbon-rich haze in exoplanet atmospheres.

This paper is organized as follows: Section~\ref{sec:2} describes the production and spectral measurement of haze analogues, along with optical constant computing and spectral modeling methods. Section~\ref{sec:3} presents the transmittance spectra, computed optical constants, and their impact on observed planetary spectra. We compare organic haze and graphite cases to evaluate their spectral differences and implications for exoplanet atmospheric characterization. Section~\ref{sec:4} summarizes the key findings and their relevance to interpreting muted or featureless spectra in carbon-rich atmospheres.

\section{Methods}\label{sec:2}
\subsection{Sample Preparation}

We performed laboratory simulations of CO$_2$-rich sub-Neptune atmospheres at 300 and 500 K using the PHAZER (Planetary Haze Research) experimental setup. The gas mixture as shown in Figure \ref{fig:1} was exposed to AC glow discharge (cold plasma) for $\sim$100 hours within the chamber. After each experiment, samples were collected as dry powders inside of the inert nitrogen glove box to avoid contamination from moisture or atmospheric oxygen. The setup and detailed experimental procedure have been described in previous studies \citep{2017HeHorst+,2018HeHorst+,2018HorstHe+,2022HeHorst+, 2024HeRadke+,2025WangYang+}. The gas compositions shown in Figure \ref{fig:1} fall well within the graphite stability regime and are predicted to form graphite based on thermochemical equilibrium modeling \citep{2013MosesLine+, 2021WoitkeHerbort+}.

\subsection{Pellet Preparation and Spectral Measurements}
We measured the transmittance spectra of the collected haze particles using the established KBr pellet method \citep{2024HeRadke+} whereby KBr is dried and powdered before mixing with the desired material. KBr is used because it has a low opacity from the UV to mid-IR. The samples and KBr were mixed at the desired ratio, and then ground to 0.7-2.2 $\mu$m using a ball-mill grinder. $\sim$200 mg of the mixture was then pressed into a pellet using a pellet die, with pressure applied in three stages: 3 tons for 2 seconds, 7 tons for 30 seconds, and 10 tons for 120 seconds for both haze and graphite pellets. The mass concentration of the organic hazes/KBr pellets is $\sim$0.3\%, which has proved to be an appropriate concentration for optical constant retrieval of organic haze materials \citep{2024HeRadke+}. For comparison, reference graphite pellets were prepared by mixing graphite powder (obtained from Sigma-Aldrich and ground to 0.7-2.2 $\mu$m) with KBr following the same procedure. Given the much stronger absorption of graphite, three graphite/KBr pellets with lower and varied mass concentrations (0.008\%, 0.0016\%, 0.03\%) were made for accurate optical constant calculation.

For both organic haze and graphite pellets, the effective optical thickness $d$ of each sample was calculated from:
\begin{equation}
    d = \frac{m}{\pi r^2 \rho}
    \label{eq1}
\end{equation}
where $m$ is the mass of haze or graphite particles in the pellet, $r$ is the pellet radius, and $\rho$ is the measured density of the samples \citep{2025WangYang+}.

All spectral measurements were performed using a Bruker Vertex 70v vacuum Fourier-transform infrared (FTIR) spectrometer at room temperature (294 K). The system was operated under vacuum conditions ($\leq$0.2~mbar) to eliminate spectral interference from atmospheric H$_2$O and CO$_2$. The instrument was equipped with two beamsplitters (quartz \& KBr) and two detectors (silicon diode \& DLaTGS) to cover the full range of 0.4–28.5~$\mu$m, encompassing the optical to mid-infrared regions relevant for exoplanetary observations. 500 scans were collected for each spectrum at a resolution of 0.5~cm$^{-1}$. A pure KBr pellet, pressed under identical conditions, served as the reference to correct potential instrument shift and scattering loss. The sample transmittance was computed as the ratio of sample to reference intensities. 
\subsection{Calculation of Optical Constants}

The imaginary part $k(\nu)$ of the complex refractive index was calculated from transmittance $T(\nu)$ via the Beer–Lambert law:
\begin{equation}
    k(\nu) = \frac{\alpha}{4\pi\nu} = \frac{1}{4\pi \nu d} \ln\left(\frac{1} {T(\nu)}\right) 
    \label{eq2}
\end{equation}

where $\alpha$ is the absorption coefficient, $d$ is the effective thickness of the sample, $\nu$ is the wavenumber in cm$^{-1}$, and $T(\nu)$ is the measured transmittance \citep{1991RoushPollack+,2012MahjoubDahoo+}. The real part $n(\nu)$ was computed using the Subtractive Kramers–Kronig (SKK) relation:
\begin{equation}
    \label{eq3}
    n(\nu) = n_0 + \frac{2(\nu^2 - \nu_0^2)}{\pi} \mathcal{P} \int_0^{\infty} \frac{\nu'k(\nu')}{(\nu'^2 - \nu^2)(\nu'^2 - \nu_0^2)} d\nu'
\end{equation}

In our analysis, the integral was divided into three regions: low-frequency (0–350~cm$^{-1}$), measured region (350–25000~cm$^{-1}$), and high-frequency (25000–$\infty$). The measured region was numerically integrated using the trapezoidal rule, while the low and high-frequency regions were extrapolated using literature values of $k(\nu)$ \citep[e.g.,][]{1984KhareSagan+,2014PapoularPapoular+}. To estimate the anchor point $n_0$ at a selected reference frequency $\nu_0$, we analyzed interference fringes in the spectra of selected film samples measured at multiple incidence angles, following the methodology of \cite{2024HeRadke+}. The fringe spacings at two incidence angles were used to determine $n_0$ at anchor frequencies. To enhance clarity in interpreting the results, a more detailed description of the numerical solution is provided in Section~\ref{sec:3}.

\subsection{Spectral Modeling with Virga and PICASO}

To simulate observational effects, we input the derived optical constants ($n$ and $k$) into Virga, a cloud code \citep{2019BatalhaMarley+,2022RooneyBatalha+}, with the custom functionality originally developed for wavelength-dependent haze optical properties \citep{2024HeRadke+}. For each haze or graphite sample, Virga was provided with the $n(\lambda)$ and $k(\lambda)$ values over the 0.5–15~$\mu$m wavelength range. The code uses PyMieScatt \citep{2018SumlinHeinson+} to compute the scattering efficiency ($Q_\mathrm{sca}$), extinction efficiency ($Q_\mathrm{ext}$), and asymmetry parameter ($g$) as functions of wavelength. These parameters were compiled into a Mie efficiency (mieff) file, formatted for use in the PICASO radiative transfer framework \citep{2019BatalhaMarley+, 2023MukherjeeBatalha+}. PICASO was used to simulate transit spectra in a limb-viewing geometry appropriate for transiting exoplanets \citep{2019BatalhaMarley+,2023AldersonWakeford+}.

Each PICASO simulation was configured with a prescribed atmospheric temperature-pressure profile, planetary surface gravity, and stellar radius representative of the modeled exoplanet. Haze-free and hazy scenarios were both explored. In hazy simulations, the haze layer was implemented using the mieff file generated by Virga. Key input parameters included the haze upper and lower pressure layer bounds and total column-integrated number density (ndz, in cm$^{-2}$). The haze layer was assumed to be vertically uniform in particle properties within its pressure bounds. For each case, the forward-modeled transit spectra were computed at a spectral resolution of R=15000. All model outputs were used to compare the spectral impact of organic haze versus graphite and assess their observational signatures across the near- and mid-infrared. Further modeling details, including gas opacities, particle sizes, and pressure–temperature profiles, are described in Section \ref{sec:3.3}.
\section{Results}\label{sec:3}
In this section, we present the key outcomes of our laboratory measurements and atmospheric modeling, aimed at quantifying the optical properties of carbon-rich hazes relevant to sub-Neptune and super-Earth atmospheres and evaluating their impact on the observed spectra.

\begin{figure*}
    \centering
    \includegraphics[width=0.85\textwidth]{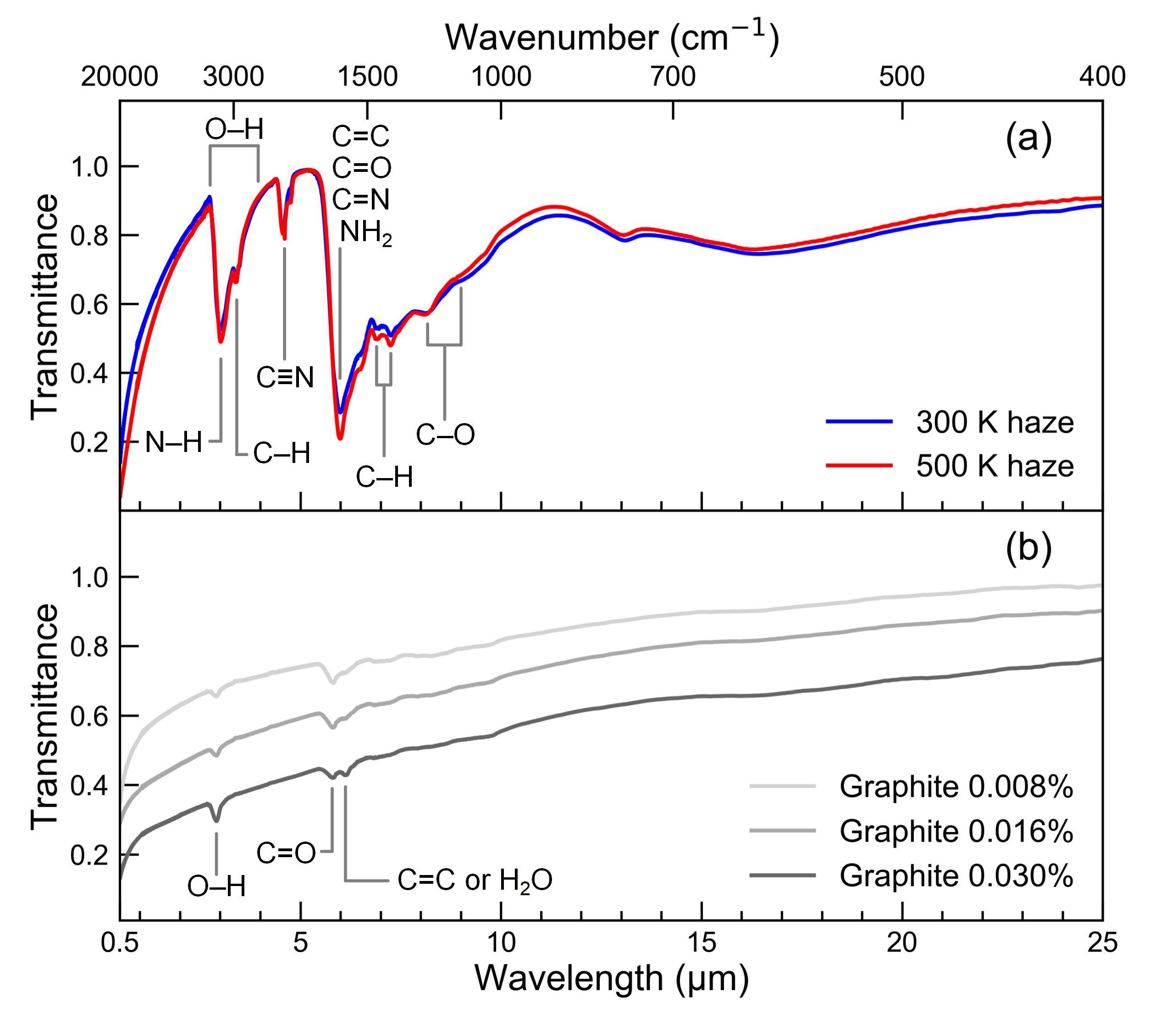}
    \caption{
    (a) Transmittance spectra of haze analogue at 300~K (blue) and 500~K (red), exhibiting rich vibrational absorption bands (e.g., N–H, C–H, C$\equiv$N, C=O). Both samples have $\sim$0.3\% concentration of organic haze and share similar spectral features, with subtle differences in band intensity reflecting temperature-dependent composition. 
    (b) Transmittance spectra of graphite pellets at three concentrations (0.008\%, 0.016\%, 0.03\%), showing smooth, broadband absorption with consistent spectral shapes (Note that graphite pellets were prepared at much lower concentrations due to the higher absorbance of graphite, and higher concentrations would result in saturated spectra). Small features near 2.9~$\mu$m, 5.8~$\mu$m, and 6.1~$\mu$m, may come from absorbed water in graphite structure and/or the by-products during the commercial production process of graphite \citep{2010MojetEbbesen+,2018BeraPragya+}, which likely arises from O-H stretching (2.9 $\mu$m), C=O stretching (5.8 $\mu$m), C=C stretching or H–O–H bending (6.1 $\mu$m) vibrations.
    } 
    \label{fig:2}
\end{figure*}

\subsection{Transmittance Spectra of Organic Haze and Graphite Samples} \label{sec:3.1}

To assess the optical behavior of the organic haze analogues and standard graphite reference material, we measure their transmittance spectra over a broad wavelength range. Figure~\ref{fig:2} presents the smoothed transmittance spectra for both classes of materials, where smoothing was performed using a Savitzky–Golay filter \citep{1964SavitzkyGolay+} to reduce high-frequency noise while preserving spectral features. The haze samples exhibit numerous well-defined molecular absorption bands, characteristic of organic hazes. Prominent features appear at 3.0~$\mu$m (N–H), 3.4~$\mu$m (C–H), 2.7–4.0~$\mu$m (O–H), 4.5~$\mu$m (C$\equiv$N), 6.0~$\mu$m (C=C, C=O, C=N, and NH$_2$), 6.9/7.2~$\mu$m (C–H), and near 8.1/9.0~$\mu$m (C–O or C-N) \citep{1991LinDaimay+}. The presence of strong O–H, C$\equiv$N, N–H, and NH$_2$ features at shorter wavelengths indicates the incorporation of polar functional groups. Although the 300 and 500~K haze samples show similar absorption profiles, subtle differences in band intensities likely reflect different reaction pathways that alter functional group abundances in the two haze samples.

In contrast, the graphite spectra (Fig.~\ref{fig:2}b) are largely featureless and exhibit a smooth, monotonic increase in absorption toward shorter wavelengths. This behavior is consistent with the delocalized electronic structure of crystalline graphite, where carbon atoms are arranged in extended sp$^2$-bonded hexagonal lattices. The associated $\pi$–$\pi^*$ electronic transitions give rise to broadband absorption without discrete vibrational signatures, yielding a continuum-like transmittance spectrum \citep[e.g.,][]{1985ConradStrauss+,2010SkulasonGaskell+,2016Draine+}. Note that weak absorption features near 2.9, 5.8, and 6.1~$\mu$m may come from absorbed water in graphite structure and/or the by-products during the commercial production process of graphite \citep{2010MojetEbbesen+,2018BeraPragya+}.
 
Importantly, all three graphite spectra exhibit nearly identical spectral shapes, with only the absolute transmittance scaling with graphite’s concentrations (0.008\%–0.03\%). This consistency confirms that the tested concentration range is well-suited for optical constant derivation—sufficiently dilute to avoid saturation effects while still preserving the material’s bulk spectral characteristics. A more detailed discussion on the applicability of our graphite samples to exoplanetary and astrophysical environments is provided later in Section \ref{sec:3.2} and Figure \ref{fig:4}.

Overall, the stark contrast in transmittance features between organic hazes and graphite materials highlights their fundamentally different optical responses. While the spectra of organic hazes reflect molecular vibrational modes of chemically diverse organic compounds in the hazes, graphite behaves as a broadband absorber with minimal molecular structure. These distinctions motivate the detailed computation of complex refractive indices ($n$ and $k$) in the following section.

\subsection{Optical Constants of Organic Hazes and Graphite}\label{sec:3.2}

\begin{figure*}
    \centering
    \includegraphics[width=0.85\textwidth]{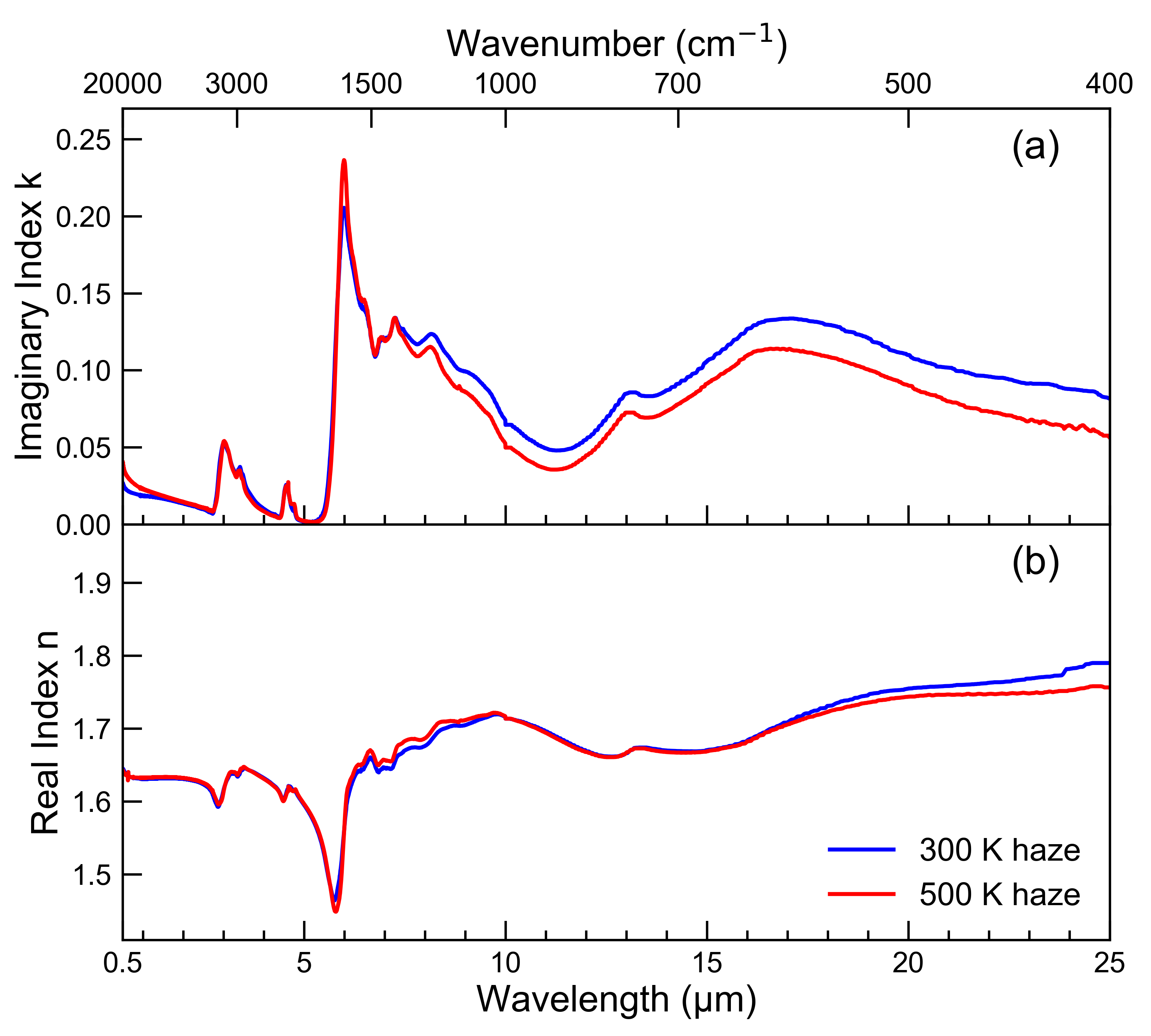}
    \caption{
    \textbf{Complex refractive indices ($n$ and $k$) of haze analogues synthesized at 300 and 500~K.}  
    (a) Extinction coefficient $k(\lambda)$ retrieved from transmittance data using the Beer–Lambert relation. Both samples exhibit multiple absorption peaks associated with vibrational modes of organic functional groups. A dominant peak near 6.2~$\mu$m is present in both spectra, with $k$ values diverging beyond 7~$\mu$m.  
    (b) Real part of the refractive index $n(\lambda)$ derived via the SKK method. A pronounced dip near 6~$\mu$m corresponds to the $k$ peak, while small but consistent differences between the 300 and 500~K samples appear across the spectrum.
    }
    \label{fig:3}
\end{figure*}

For the organic haze samples, the extinction coefficient $k(\nu)$ was derived from the Beer–Lambert law (Equation \ref{eq2}) using the effective sample thicknesses, determined to be $3.16\times10^{-4}$~cm for the 300~K haze and $2.91\times10^{-4}$~cm for the 500~K haze. The corresponding real part of the refractive index, $n(\nu)$, was then obtained via the SKK integral transformation (Equation \ref{eq3}). To ensure numerical stability and robustness, we computed $n(\nu)$ over 31 evenly spaced reference frequencies $\nu_0$ from 12500 to 14000 cm$^{-1}$ with $n_0$ determined using the fringe methods. The final $n$ spectrum was taken as the median of these solutions with the final anchor point $n_0 = 1.686$ at $\nu_0 = 13100$~cm$^{-1}$ for the 300 K sample and $n_0 = 1.631$ at $\nu_0 = 13100$~cm$^{-1}$ for the 500 K sample. Figure 3 displays the retrieved optical constants ($k$ and $n$) for the 300 K (blue) and 500 K (red) haze analogues. In the extinction coefficient panel (Fig~\ref{fig:3}a), both samples show a prominent absorption peak near 6.0 $\mu$m, accompanied by several narrower features across the 3–10 $\mu$m region, corresponding to specific vibrational modes of different organic functional groups in the haze samples. Importantly, a clear divergence appears between the two haze samples beyond 7 $\mu$m, where the 300 K haze consistently exhibits higher $k$ values than the 500 K sample. In the refractive index profiles (Fig~\ref{fig:3}b), both samples show relatively smooth variations, with a distinct dip near 6 $\mu$m corresponding to the major $k$ peak. The two n curves are generally similar in shape but exhibit subtle offsets throughout the spectrum. These distinctions indicate compositional differences arising from the experimental condition.

We derived the optical constants of commercial graphite from our measured transmittance spectra using two methods. The first method utilized the conventional single-pellet Beer–Lambert law (Equation \ref{eq2}) to derive the extinction coefficient $k(\nu)$ from each of the three graphite–KBr pellets (0.03\%, 0.016\%, and 0.008\%), and the three k values were averaged to yield a representative $k(\nu)$ spectrum. The second method applied a multi-concentration linear fitting procedure to the same set of pellets, corresponding to effective graphite thicknesses of 19.89, 10.55, and 5.12~$\mu$m, respectively. Rather than treating each pellet separately, we simultaneously modeled their transmittance spectra using the relation:
\begin{equation}
    -\ln T(\nu) = \alpha(\nu) \cdot d + b(\nu)
    \label{eq4}
\end{equation}
In this formulation, $\alpha(\nu)$ represents the intrinsic absorption coefficient of graphite and $b(\nu)$ is a wavelength-dependent baseline term accounting for light scattering, reflection losses, or detector drift. A point-by-point linear regression was performed across all three datasets to extract $\alpha(\nu)$, from which the extinction coefficient $k(\nu)$ was calculated using standard relations (Equation \ref{eq2}). 

Using the SKK relation (Equation \ref{eq3}), we calculated the n($\nu$) values at 21 evenly spaced reference frequencies $\nu_0$ from 5000 to 15000 cm$^{-1}$ and took the median as the final n values. For the average method, the transformation was anchored at $\nu_0 = 9000$~cm$^{-1}$ with $n_0 = 2.8$; for the fitting method, it was anchored at $\nu_0 = 7000$~cm$^{-1}$ with $n_0 = 2.8$, following \cite{2014PapoularPapoular+}.

\begin{figure*}
    \centering
    \includegraphics[width=0.85\textwidth]{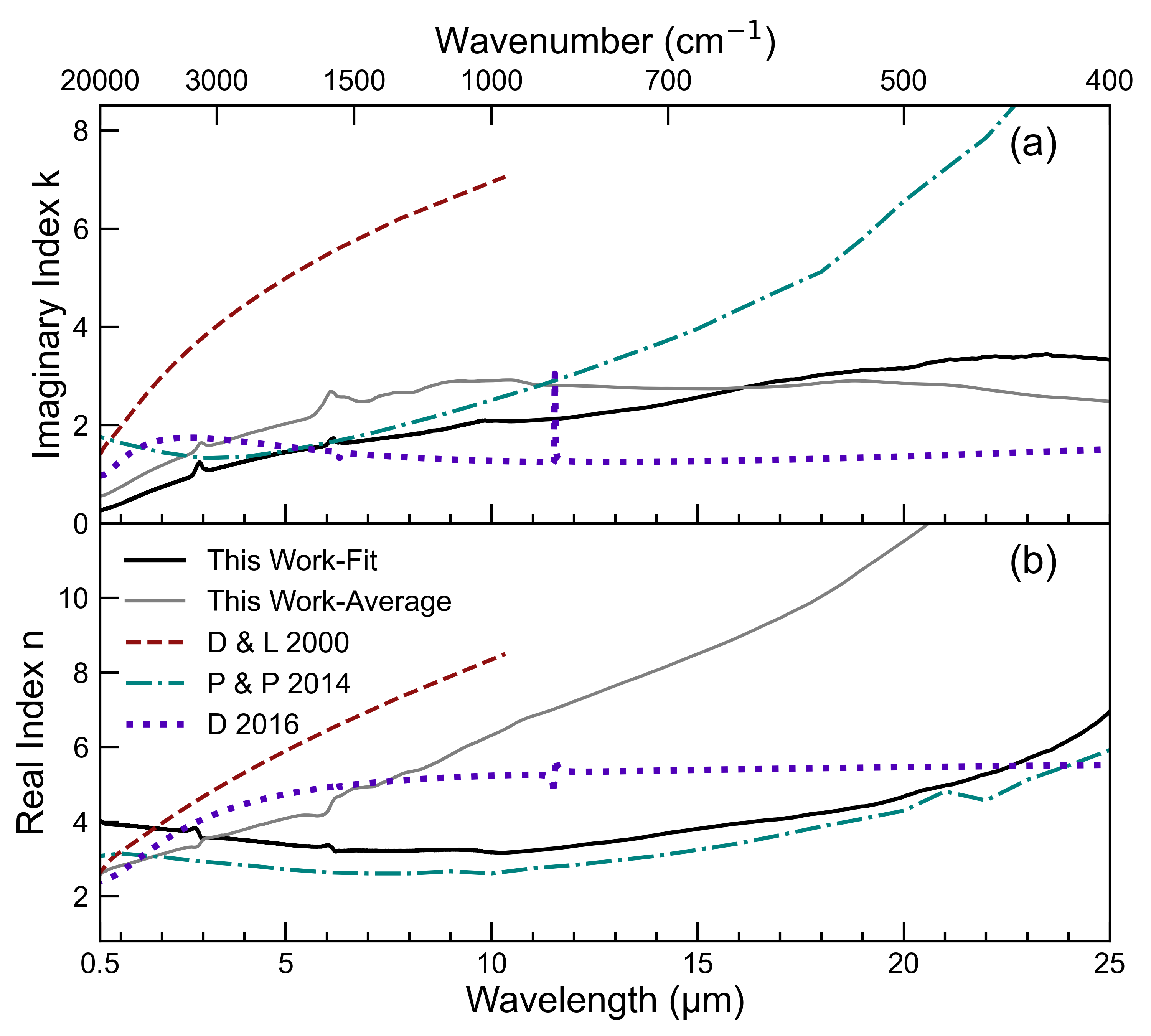}
    \caption{
    \textbf{Complex refractive indices ($n$ and $k$) of graphite compared with previous studies.}  
    (a) Extinction coefficient $k(\lambda)$ derived using two approaches: multi-concentration fitting (black) and single-pellet averaging (gray). Prior datasets from \cite{2000DjurisicLi+,2014PapoularPapoular+,2016Draine+} (D \& L 2000 brown, P \& P 2014 turquoise, D 2016 indigo) are overlaid for comparison, illustrating the wide variation in reported graphite optical constants.  
    (b) Real part of the refractive index $n(\lambda)$ retrieved via the SKK method. The fit-based result (black) shows better alignment with \cite{2014PapoularPapoular+}, while the averaged result (gray) displays a steeper rise. The discrepancies among all datasets reflect differences in material preparation, spectral methods, and analytical assumptions, underscoring the difficulty in constraining graphite's complex optical properties.
    }
    \label{fig:4}
\end{figure*}

Figure~\ref{fig:4} compares our computed complex refractive indices for commercial graphite using the two approaches—the averaged method (gray) and the fitting method (black)—alongside three representative literature datasets. In both methods, the extinction coefficient $k(\lambda)$ rises toward longer wavelengths, consistent with $\pi \rightarrow \pi^*$ electronic transitions in delocalized sp$^2$ carbon structures. Notably, the fitting method produces a smoother and more monotonic $k$ spectrum, particularly in the 5–10~$\mu$m range. The fitted method corrects potential systematic errors like light scattering, reflection losses, or detector drift, and yields a reliable and physically meaningful refractive index. In contrast, the averaged $k$ values overestimate absorption in this region, propagating into an inflated real refractive index $n(\lambda)$ via the SKK relation. Therefore, the data calculated from the fitted method is used for the spectral modeling (Section \ref{sec:3.3}). Note that we calculated the optical constants for the organic hazes using one concentration at 0.3\%, as our previous study \citep{2024HeRadke+} has demonstrated that this concentration yields results comparable to those obtained using the fitted methods for similar organic haze materials.

Our fitted $k$ values lie between those reported by \cite{2014PapoularPapoular+, 2016Draine+} across most of the spectral range, while the \cite{2016Draine+} dataset exhibits significantly higher $k$ values and a steep rise below 10~$\mu$m. This indicates a stronger absorption trend in the \cite{2000DjurisicLi+}'s results, likely influenced by methodological differences. For the real refractive index $n$, our fitted $n$ values fall between the lower bound defined by \cite{2014PapoularPapoular+} and the higher bound of \cite{2016Draine+}, closely following the shape and slope of \cite{2014PapoularPapoular+}'s curve throughout the mid-infrared. In contrast, \cite{2000DjurisicLi+} reports much higher $n$ values with a sharp increase beyond 10~$\mu$m, diverging from the more moderate trend seen in our computed values.

The considerable variation in reported graphite optical constants underscores the persistent challenges in accurately determining graphite's optical properties. As stated in \cite{2016Draine+}, optical constants are derived through a variety of experimental techniques, including reflectivity measurements, electron energy loss spectroscopy (EELS), and photoemission at X-ray energies \citep[e.g.,][]{1991Palik+,2016Draine+}. Each method probes different energy regimes and may introduce systematic uncertainties depending on factors such as sample orientation, grain size, degree of anisotropy, and surface treatment. In this study, we derived the optical constants of graphite using the measured transmittance spectra of small, randomly-oriented graphite particles (0.7-2.2 $\mu$m) dispersed in transparent KBr medium. The states of graphite particles in the measurement setting are more representative to graphite hazes in atmospheres or graphite grains in molecular clouds and the interstellar medium, compared to other methods such as reflectance spectroscopy on polished bulk samples, electron energy loss spectroscopy (EELS), or thin-film ellipsometry. Moreover, our transmission measurements on dispersed particles better reflect the geometry and conditions relevant to transit or absorption spectroscopy. Therefore, the graphite optical constants derived here are more suitable for interpreting astronomical observations.

\subsection{Radiative Transfer Modeling with PICASO}\label{sec:3.3}
We employed Virga \citep{2019BatalhaMarley+,2022RooneyBatalha+} and PICASO \citep{2019BatalhaMarley+, 2023MukherjeeBatalha+} to evaluate the radiative impact of the measured optical constants under conditions matching our laboratory experiments. The pressure–temperature ($P$–$T$) profile used in the simulations is isothermal at 500~K, spanning 59 logarithmically spaced pressure layers from 10$^{-10}$ to 10$^2$~bar, consistent with the environment used in our haze production experiments. The atmospheric gas composition was consistent with the initial gas mixture used in the 500 K laboratory experiments (Figure \ref{fig:1}), with volume mixing ratios of CO$_2$, CH$_4$, N$_2$, H$_2$O, He, and H$_2$ fixed vertically. These abundances were derived from CEA equilibrium modeling at 2000$\times$ solar metallicity and a solar C/O ratio, and used directly in radiative transfer. For the opacities of the species, we use the V3 Zenodo Opacity Database \citep{2025BatalhaFreedman+}, with the most important gas phase absorbers being H$_2$O \citep{Polyansky_2018}, CH$_4$ \citep{Yurchenko_2013,Yurchenko_2014}, and CO$_2$ \citep{Huang_2014} given our model atmosphere compositions.

The particle size distribution followed a log-normal form with radii spanning 20–200 nm \citep{2024HeRadke+}. Organic haze layers were implemented with a constant number density of 50 cm$^{-3}$ between 0.1 bar and 1 nbar, corresponding to a total column density of $1.5\times10^{10}$ cm$^{-2}$. A representative sub-Neptune exoplanet, GJ1214b, was chosen as the benchmark due to its well-characterized transit spectra and hypothesized haze-rich atmosphere. Planetary parameters adopted were: mass 8.17~$M_\oplus$, radius 2.742~$R_\oplus$, and equilibrium temperature $\sim$596 K from \citet{2021CloutierCharbonneau+}; stellar radius $0.2162$~$R_\odot$ from \citet{2024MahajanEastman+}.

\begin{figure*}[ht]
    \centering
    \includegraphics[width=\linewidth]{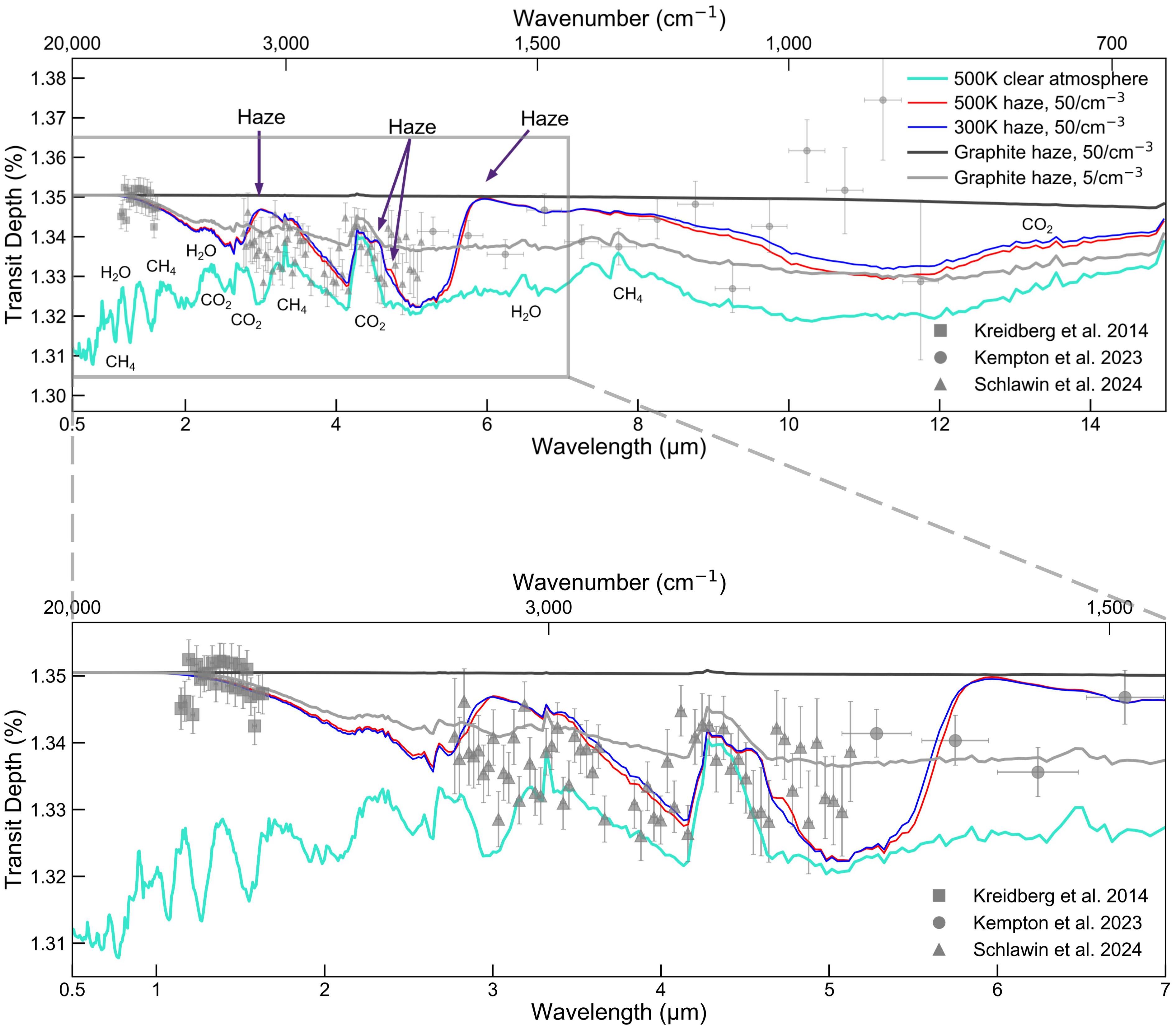}
    \caption{
    \textbf{Modeled transit spectra of GJ~1214b including haze and graphite scenarios.}
    Planet mass 8.17~$M_\oplus$, radius 2.742~$R_\oplus$, and equilibrium temperature $\sim$596~K are adopted from \cite{2021CloutierCharbonneau+}. The stellar radius, $0.2162~R_\odot$, are adopted from \cite{2024MahajanEastman+}. Simulations include a cloud-free atmosphere (cyan), haze analogs produced at 300~K (blue) and 500~K (red), graphite-based aerosols (dark gray), and the graphite-based aerosols using 10\% of the number density (5~cm$^{-3}$ volumetric, light gray) overlaid with observational data from \cite{2014KreidbergBean+,2023KemptonZhang+}, and \cite{2025SchlawinOhno+}. The upper panel shows the full wavelength range (0.5-15~$\mu$m), while the lower panel provides a magnified view of the visible to near-infrared region (0.5–7~$\mu$m), emphasizing haze-induced modulations and absorption features. Hazy scenarios suppress molecular absorption bands and introduce new spectral signatures (e.g., at 3.0, 4.5, 4.7, and 6.0~$\mu$m). Graphite shows extreme flattening due to its broadband opacity.
    }
    \label{fig:5}
\end{figure*}
Figure~\ref{fig:5} presents the synthetic spectra and the observational data for GJ~1214b from \cite{2014KreidbergBean+, 2023KemptonZhang+}, and \cite{2025SchlawinOhno+} for comparison. To facilitate visual comparison of model features and different datasets, we apply an offset of 0.012\% in transit depth to the data from \cite{2025SchlawinOhno+}. The synthetic spectrum of the clear atmosphere shows pronounced absorption features from key gases such as H\textsubscript{2}O, CH\textsubscript{4}, and CO\textsubscript{2}. In contrast, the hazy scenarios exhibit substantial muting of molecular features across the spectrum, particularly below 2~$\mu$m, where the spectra become nearly flat. Between 3 and 7~$\mu$m, distinct differences emerge: both organic haze models suppress gas features to varying degrees, while simultaneously introducing new spectral features at 3.0,
4.5, 4.7, and 6.0 $\mu$m—attributable to organic functional groups in the hazes. Across all four features, our 500 K haze consistently shows stronger absorption than the 300 K case, suggesting that elevated formation temperatures promote haze species with greater infrared activity. 

Notably, the synthetic spectrum for graphite at a number density (50 cm$^{-3}$)  is nearly featureless, with atmospheric gas absorption features almost entirely suppressed and a consistent uplift in transit depth (0.025\%). This flat, absorbing behavior highlights the broadband opacity of graphite and suggests that if such material dominates the aerosol composition, spectral signatures could be entirely masked. Even at a much lower number density (5 cm$^{-3}$), graphite introduces stronger masking effects than the organic hazes at 10 times the number density (50 cm$^{-3}$), allowing only a few molecular features to emerge.

A comparison between our synthetic spectra and GJ~1214b observations shows that organic haze models capture several qualitative features present in the data. The organic haze models are broadly consistent with the flat near-IR spectrum reported by \cite{2014KreidbergBean+} in the near IR region(1.1–1.6~$\mu$m, HST/WFC3), supporting the presence of high-opacity hazes in this region. In the 2.8–5.1~$\mu$m region reported by \cite{2025SchlawinOhno+}(JWST/NIRSpec), some haze-specific absorption features—especially near 4.7~$\mu$m—partially align with our simulations, though the observed band intensities differ, suggesting possible compositional or abundance variations. For the mid-IR data (5.2–11.7~$\mu$m), \citep{2023KemptonZhang+}, overall trends are broadly consistent but exhibit larger uncertainties. \cite{2025SchlawinOhno+}'s spectrum shows strong absorption near 4.3~$\mu$m, matching the CO$_2$ feature in our clear and hazy model and indicating CO$_2$ present in the atmosphere on GJ~1214b.

We use GJ 1214b as an example to demonstrate the effect of different haze optical constants on the observed spectra. Our aim is not to reproduce the observed spectrum in detail, but rather to isolate and demonstrate the spectral impact of haze composition under otherwise fixed model conditions. These haze particles are not necessarily present in GJ~1214b's atmosphere, but their characteristic absorption features may serve as diagnostic signatures in future observations of carbon-rich exoplanet atmospheres.

Moreover, the near-flat spectrum with graphite haze indicates that graphite is not the dominant haze in the atmosphere of GJ 1214b. While laboratory results show that graphite does not form under atmospheric conditions of $500\:\mathrm{K}/300\: \mathrm{K}$ and 5 mbar \citep{2025WangYang+}, graphite could form in deep, hot regions of the atmosphere and be lofted into observable altitudes. Such a spectrally flat, broadband absorber could obscure molecular features and mimic the effect of a planetary surface on transit geometry. In this scenario, the top of the graphite haze layer may be interpreted as the true planetary radius, leading to an overestimated radius and, consequently, an underestimated bulk density. This mechanism offers a plausible explanation for super-puff planets \citep[e.g.,][]{2020GaoZhang+,2021OhnoTanaka+}, whose anomalously low densities may be apparent rather than intrinsic.

\section{Conclusion}\label{sec:4}
In summary, our results reveal distinct spectral differences between haze and graphite. The organic haze exhibits identifiable absorption features from various functional groups, while graphite displays featureless, broadband absorption that strongly masks molecular features in simulated transit spectra.

This study provides the first set of optical constants for organic hazes formed in CO$_2$-rich, high-metallicity environments, filling a key gap in existing aerosol datasets. The results are expected to be representative for organic hazes formed in CO$_2$-rich sub-Neptune atmospheres within the graphite-stability regime, spanning from 2000-10000× metallicity with C/O ratios from 0.5-0.85 as predicted by thermochemical equilibrium modeling (CEA). Our graphite data, derived via spectra of randomly oriented and dispersed sub-micron particles, better represent the particulate state of atmospheric or interstellar graphite than measurements based on bulk or film samples. Compared to previous datasets based on reflectance, ellipsometry, or EELS measurements on bulk or oriented samples, our approach is more directly applicable to astronomical observations relying on transmission geometry.

Using the derived optical constants, we simulated transit spectra of a representative sub-Neptune atmosphere using  Virga and PICASO. The haze models reproduce several spectral features that are potentially consistent with those observed in GJ~1214b, suggesting the widespread presence of high-opacity photochemical hazes. Additionally, the distinctive haze absorption features (e.g., near 3.0, 4.5, 4.7, and 6.0 $\mu$m) provide potential diagnostic markers for future observations aiming to identify haze compositions in carbon-rich atmospheres.

Meanwhile, the graphite model yields a nearly flat spectrum, effectively masking molecular features and mimicking the spectral signature of a solid planetary surface, offering a plausible explanation for the anomalously low densities of super-puff exoplanets like Kepler 51d. 

Overall, our study presents an experimentally-grounded optical dataset for both haze and graphite relevant to CO$_2$-rich sub-Neptune atmospheres, enabling more accurate spectral retrievals and informing the interpretation of muted and featureless exoplanet spectra. A full model parameter study is required to understand the effects of the organic hazes on exoplanet observations.

\begin{acknowledgments}
The authors thank the anonymous referee for their valuable comments that enhanced the presentation of this paper. The authors gratefully acknowledge the supports from the National Natural Science Foundation of China (42475132) and the U.S. National Science Foundation (2206245). S.E.M. is supported by NASA through the NASA Hubble Fellowship grant HST-HF2-51563 awarded by the Space Telescope Science Institute, which is operated by the Association of Universities for Research in Astronomy, Inc., for NASA, under contract NAS5-26555. V.Vuitton acknowledges support from the French National Research Agency in the framework of the “Investissements d’Avenir” program (ANR-15-IDEX-02), through the funding of the Origin of Life project of the Université Grenoble Alpes and the French Space Agency (CNES) under their “Exobiologie, Exoplanètes et Protection Planétaire” program.
\end{acknowledgments}

\bibliography{bib}{}
\bibliographystyle{aasjournalv7}

\end{document}